%% file: main.tex
\documentclass[10pt,conference]{IEEEtran}
\IEEEoverridecommandlockouts

\usepackage{cite}
\usepackage{amsmath,amssymb,amsfonts}
\usepackage{graphicx}
\usepackage{textcomp}
\usepackage{xcolor}
\def\BibTeX{{\rm B\kern-.05em{\sc i\kern-.025em b}\kern-.08em
    T\kern-.1667em\lower.7ex\hbox{E}\kern-.125emX}}

\usepackage[linesnumbered,ruled,vlined,noend]{algorithm2e}
\usepackage{url}
\usepackage{subcaption}
\usepackage{caption}
\usepackage{quantikz}
\usepackage{booktabs}
\usepackage{array}
\usepackage{multirow}
\usepackage[colorlinks=true, linkcolor=blue, citecolor=green, urlcolor=magenta]{hyperref}
\usepackage[most]{tcolorbox}
\usepackage{enumitem}
\usepackage{footnote}
\setlist[enumerate,1]{label=\arabic*., leftmargin=1em}

\begin{document}

\input{paper/macro}

\title{CLASS: A Controller-Centric Layout Synthesizer for Dynamic Quantum Circuits}

\author{
    \IEEEauthorblockN{
      Yu Chen\textsuperscript{1,2,\ref{equal_contrib}}, Yilun Zhao\textsuperscript{1,2,\ref{equal_contrib}}, Bing Li\textsuperscript{3}, He Li\textsuperscript{4}, Mengdi Wang\textsuperscript{1}, Yinhe Han\textsuperscript{1}, and Ying Wang\textsuperscript{1,\ref{corresponding_author}}
    }
    \IEEEauthorblockA{    
        \textit{Research Center for Intelligent Computing Systems, Institute of Computing Technology, Chinese Academy of Sciences}\textsuperscript{1} \\
        \textit{University of Chinese Academy of Sciences}\textsuperscript{2},
        \textit{Institute of Microelectronics, Chinese Academy of Sciences}\textsuperscript{3},
        \textit{Southeast University}\textsuperscript{4} \\
        Emails: \{chenyu21b,zhaoyilun22b,wangmengdi,yinhes,wangying2009\}@ict.ac.cn,libing2024@ime.ac.cn,helix@seu.edu.cn \vspace{-12pt}
    }
}

\maketitle

\begingroup\renewcommand\thefootnote{*}
\footnotetext{\label{corresponding_author}Corresponding author.}
\endgroup

\begingroup\renewcommand\thefootnote{$\dagger$}
\footnotetext{\label{equal_contrib}Equal contribution.}
\endgroup

\input{paper/0-abstract}


\input{paper/0-intro-v2}

\input{paper/1-background}

\input{paper/2-motivation}

\input{paper/3-design-0}

\input{paper/3-design-1}

\input{paper/4-evaluation}

\input{paper/5-discussion}

\section{Conclusion and Future Work}

This work addresses the challenges posed by inter-controller communication delays in the layout synthesis of dynamic quantum circuits (DQCs).
For DQCs without connectivity constraints, we model the problem as a minimum-cut task in an undirected hypergraph and solve it using an efficient heuristic approach. 
This solution provides the initial placement for DQCs with connectivity constraints. 
To mitigate the impact of SWAP insertions during gate scheduling, 
we enhance existing schedulers with an effective mechanism.
Our evaluations demonstrate that our synthesizer achieves up to a 100\% reduction in inter-controller communications, with only a \(\sim\)2\% increase in additional CNOTs. 

\section*{Acknowledgments}

We thank Dr. Xiang Fu for his valuable feedback and insightful discussions. This work is supported in part by National Natural Science Foundation of China (NSFC) (62025404, 62222411, 62304037), National Key Research and Development Program of China (2023YFB4404400), and the Natural Science Foundation of Jiangsu Province under Grant BK20230828.

\renewcommand{\IEEEbibitemsep}{0pt plus 0.5pt}
\makeatletter
\IEEEtriggercmd{\reset@font\normalfont\fontsize{7.9pt}{8.40pt}\selectfont}
\makeatother
\IEEEtriggeratref{1}
\begingroup
\scriptsize 
\bibliographystyle{IEEEtran}
\bibliography{references}
\endgroup

\end{document}

%% file: paper/macro.tex
\newcommand{\bln}[1]{\textcolor{purple}{#1}}
\newcommand{\red}[1]{\textcolor{red}{#1}}

\newcommand{\name}{CLASS}

\newcommand{\shortparadigm}{SCME}

\newcommand{\squishlist}{
   \begin{list}{$\bullet$}
    {
    \setlength{\itemsep}{0pt}      \setlength{\parsep}{0pt}
      \setlength{\topsep}{3pt}       \setlength{\partopsep}{0pt}
      \setlength{\listparindent}{-2pt}
      \setlength{\itemindent}{-5pt}
      \setlength{\leftmargin}{1em} \setlength{\labelwidth}{0em}
      \setlength{\labelsep}{0.5em} } }

\newcommand{\squishend}{
    \end{list}  }

\newcommand*\circled[1]{\tikz[baseline=(char.base)]{
  \node[shape=circle,draw,fill=black,text=white,font=\bf,inner sep=0.5pt] (char)
  {\scriptsize#1};
}}

\newcommand*\circledwhite[1]{\tikz[baseline=(char.base)]{
  \node[shape=circle,draw,fill=white,text=black,font=\bf,inner sep=0.5pt] (char)
  {\scriptsize#1};
}}

\newcommand{\etal}{et al.}
\newcommand{\ie}{i.e.}
\newcommand{\eg}{e.g.}
\newcommand{\etc}{etc.}

\newcommand{\putsec}[2]{\vspace{-0.1in}\section{#2}\label{sec:#1}\vspace{-0.05in}}
\newcommand{\putssec}[2]{\vspace{-0.0in}\subsection{#2}\label{ssec:#1}\vspace{-0.0in}}
\newcommand{\putsssec}[2]{\vspace{-0.0in}\subsubsection{#2}\label{sssec:#1}\vspace{-0.0in}}
\newcommand{\putsssecX}[1]{\vspace{0.0in}\noindent\textbf{#1:}}

\newcommand{\figref}[1]{Fig.~\ref{fig:#1}}
\newcommand{\eqnref}[1]{Eq.~\ref{eq:#1}}
\newcommand{\tabref}[1]{Table~\ref{tab:#1}}
\newcommand{\secref}[1]{Section~\ref{sec:#1}}
\newcommand{\ssecref}[1]{Section~\ref{ssec:#1}}
\newcommand{\sssecref}[1]{Section~\ref{sssec:#1}}

\newcommand{\COMM}[1]{#1}
\newcommand{\yilun}[1]{\COMM{\textcolor{cyan}{\sf\bfseries Yilun: #1}}}
\newcommand{\blc}[1]{\COMM{\textcolor{blue}{\sf\bfseries[BL: #1]}}}

\renewcommand{\COMM}[1]{}

\newcommand{\TODO}[1]{#1}
\newcommand{\todo}[1]{\TODO{{\color{red}\sf\bfseries [#1]}}}

\definecolor{colorcontent}{HTML}{FBE59E}
\definecolor{mygreen}{HTML}{85B243}
\definecolor{colortitle}{HTML}{CEE55D}

\newtcolorbox{insightbox}[1]{
  colback=colorcontent,
  colframe=black,
  coltitle=black,
  fonttitle=\bfseries,
  title=#1,
  sharp corners,
  boxrule=1pt,
  enhanced,
  top=1mm,
  bottom=1mm,
  left=1mm,
  right=1mm,
  titlerule=0.2mm,
  title filled,
  colbacktitle=colortitle,
  before skip=10pt,
  after skip=10pt
}

%% file: paper/0-abstract.tex
\begin{abstract}

Layout Synthesis for Quantum Computing (LSQC) is a critical component of quantum design tools.
Traditional LSQC studies primarily focus on optimizing for reduced circuit depth by adopting a device-centric design methodology.
However, these approaches overlook the impact of classical processing and communication time, thereby being insufficient for Dynamic Quantum Circuits (DQC).

To address this, we introduce \name{}, a controller-centric layout synthesizer designed to reduce inter-controller communication latency in a distributed control system.
It consists of a two-stage framework featuring a hypergraph-based modeling and a heuristic-based graph partitioning algorithm.
Evaluations demonstrate that \name{} effectively reduces communication latency by up to 100\% with only a 2.10\% average increase in the number of additional operations.

\end{abstract}

%% file: paper/0-intro-v2.tex
\section{Introduction}

Similar to the design of classical circuits and systems, realizing conceptual quantum algorithms on actual devices requires a multitude of complex design tasks~\cite{wille2022basis_design_tool}.
One of the most challenging design tasks is
\emph{Qubit mapping}~\cite{li2019tackling}, or \emph{Layout Synthesis for Quantum Computing} (LSQC)~\cite{tan2020OLSQ}.

Prior LSQC studies have predominantly followed a \emph{device-centric} design methodology,
with a primary focus on minimizing the execution time of quantum operations on quantum devices.
As a result, substantial effort has been devoted to reducing the number of SWAP gates induced by topological constraints of physical qubits.
~\cite{lin2023scalable_olsq,tan2020OLSQ,li2019tackling,wille2023mqt_qmap,zhang2021timeoptimal_qubit_mapping,park2022fast_qubit_map_nisq,wille2019mapping_ibm_qx,tang2024alpharouter_lsqc,yang2024quantum_lsqc_incre_sat}.

While the device-centric methodology is effective for \emph{static} quantum circuits, it falls short in optimizing for \emph{Dynamic} Quantum Circuits (DQC), a promising paradigm essential for various quantum experiments~\cite{dqc_corcoles2021exploiting,baumer2024dqc_qft,IBM_dqc,vazquez2024scaling_qc_with_dqc,shirizly2024rb_dqc}.
The primary reason is that a quantum program's execution time depends on the instruction processing time on Quantum Control Processors (QCPs)~\cite{fu2017quma}.
In static circuits, this processing time is comparable to the duration of quantum device operations (cf. Sec.~\ref{sec:problem_and_motivation}), making the device-centric approach reasonable.
In contrast, DQCs often involve frequent mid-circuit measurements and feedforward operations~\cite{dqc_corcoles2021exploiting,baumer2024dqc_qft}, introducing additional latency that cannot be captured by on-device quantum operations.
Therefore, a controller-centric design methodology becomes essential to accurately account for the total execution latency on QCPs.

\subsection{Controller-Centric LSQC Challenge}

Achieving quantum advantage requires scaling up to thousands or even millions of qubits~\cite{IBM_roadmap_latest,GoogleRoadmap}.
To support such large-scale quantum systems, it is natural to adopt a distributed control architecture composed of numerous QCPs.

When executing DQCs on distributed control systems, inter-controller communication becomes a critical performance bottleneck.
Specifically, feedforward operations on certain qubits may depend on measurement outcomes from qubits managed by different QCPs.
As a result, inter-controller communication is required to exchange measurement results or branching flags---an expensive operation in quantum computing due to the limited coherence time of qubits.
Moreover, the latency of such communication inevitably increases with system scale.

To mitigate this overhead and preserve execution fidelity, two complementary approaches can be considered.
On the hardware side, latency can be reduced through improved communication protocols and specialized interconnects.
On the software side, the mapping of logical qubits to controllers---determined by the layout synthesizer---plays a crucial role.
For instance, if all qubits involved in a feedforward operation are assigned to the same controller, the operation can be executed locally, eliminating inter-controller communication entirely.

This leads to a clear design objective for controller-centric layout synthesis:

\begin{insightbox}{Design Goal}
For each measurement-feedforward operation, the involved qubits should be mapped to a set of controllers that minimizes inter-controller communication latency.
\end{insightbox}

Unfortunately, existing LSQC solutions are designed without considering the above objective, making their modeling approaches difficult to adapt to this emerging problem.
Therefore, it remains an open challenge to design a controller-centric layout synthesizer to optimize for lower inter-controller communication latency.

\subsection{Contributions}

In this paper, we present \name{}, a systematic approach to address the emerging LSQC challenge in distributed control systems.
Our key contributions are as follows:

\begin{enumerate}

\item \textbf{A controller-centric design methodology.}
Through detailed analysis, we identify the instruction processing time of quantum control systems as the dominant factor affecting program execution time.
This insight reveals that considering only on-device operation latency is insufficient when designing tools for quantum computing.

\item \textbf{\name{}: a \underline{C}ontroller-centric \underline{LA}yout \underline{S}ynthe\underline{S}izer.}
\name{} reformulates the LSQC problem as a hypergraph partitioning task, enabling a concise and modular algorithmic framework that can be seamlessly integrated into existing layout synthesis pipelines.

\item \textbf{Evaluation of \name{}.}
We evaluate \name{} across a variety of DQC benchmarks, demonstrating its effectiveness in reducing inter-controller communication latency.
Compared to existing synthesizers, \name{} achieves an average 48.45\% reduction in inter-controller communication hops, with only a 2.10\% average increase in additional operations.
Our implementation is open-source\footnote{\url{https://github.com/Zhaoyilunnn/dqc-map}}.

\end{enumerate}

%% file: paper/1-background.tex
\section{Background and Motivation}

In this section, we commence with a concise overview of existing layout synthesizers.
Next, we introduce the concepts of dynamic quantum circuits and quantum control systems, providing examples to aid readers in understanding the specific problem addressed in this paper.
Basic concepts of quantum computing theory have been omitted for brevity; readers seeking foundational knowledge may refer to textbooks such as Ref.~\cite{nielsen2010qc_and_qi} for an in-depth introduction.

\subsection{Layout Synthesis for Quantum Computing}

A \emph{quantum circuit}\footnote{We use quantum circuit and quantum program interchangeably throughout this paper.} is a graphical representation of a quantum algorithm, consisting of a sequence of quantum gates or operations applied to logical qubits\footnote{The term ``logical" in this work does not refer to quantum error correction.}.
Implementing two-qubit gates requires physical connectivity, but the limited connectivity of most quantum devices often renders quantum circuits non-executable on such hardware.
To address this challenge, LSQC has become a critical component of modern quantum design tools, which typically consists of two stages:
(1) generating an \emph{initial placement}, which maps logical qubits to physical qubits, and
(2) producing a \emph{gate schedule}, which determines where and when to apply SWAP operations to enable the circuit’s execution on the target device.
Previous studies on LSQC can be categorized into three main approaches.
The first employs heuristic search strategies, modeling the quantum circuit as a directed acyclic graph (DAG) and using breadth-first search-like methods for SWAP insertion~\cite{li2019tackling,tannu2019not,fu2023effective_and_efficient_mapper}.
The second formulates LSQC as a mathematical optimization problem and solves it using specialized solvers~\cite{molavi2022maxsat,yang2024quantum_lsqc_incre_sat,tan2020OLSQ,huang2024smt_silicon_qc,shaik2023optimal_classical_plan}.
The third leverages machine learning techniques, such as reinforcement learning, to tackle LSQC~\cite{tang2024alpharouter_lsqc,quetschlich2023compiler_opt_qc_rl}.
Among these, heuristic methods are the most widely adopted due to their efficiency and stability.
For example, SABRE~\cite{li2019tackling}, one of the most prominent heuristic layout synthesizers, has been integrated into IBM’s Qiskit framework~\cite{javadi2024quantum_with_qiskit}, the most widely used quantum computing software.
It is important to note that all the aforementioned approaches primarily focus on \emph{static} quantum circuits without measurement feedforward operations, making them complementary to our study.

\subsection{Dynamic Quantum Circuits}

\yilun{emphasize the importance of DQC}

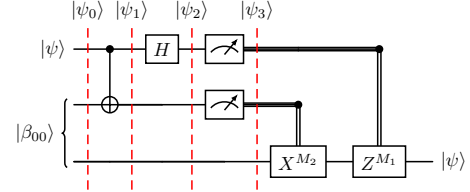
\begin{figure}[h]
    \centering
    \scalebox{0.72}{
    \begin{quantikz}
        \lstick{$|\psi\rangle$} \slice{$|\psi_0\rangle$}
        & \ctrl{1}\slice{$|\psi_1\rangle$} & \gate{H}\slice{$|\psi_2\rangle$}   & \meter{}\slice{$|\psi_3\rangle$}  & \cw             & \cwbend{2} \\
        \lstick[wires=2]{$|\beta_{00}\rangle$}
        & \targ{}  & \qw        & \meter{}  & \cwbend{1} \\
        & \qw      & \qw        & \qw       & \gate{X^{M_2}}  & \gate{Z^{M_1}} & \rstick{$|\psi\rangle$}
    \end{quantikz}
    }
    \caption{Quantum circuit for teleporting a qubit.}
    \label{fig:quantum_tele}
\end{figure}

\begin{equation}
    \scriptsize
    \begin{aligned}
    |\psi_1\rangle &= \frac{1}{\sqrt{2}} \Bigl[\alpha|0\rangle\left(|00\rangle + |11\rangle\right) + \beta|1\rangle\left(|10\rangle + |01\rangle\right)\Bigr],   \\
    |\psi_2\rangle &= \frac{1}{2} \Bigl[\alpha\left(|0\rangle + |1\rangle\right)\left(|00\rangle + |11\rangle\right) + \beta\left(|0\rangle - |1\rangle\right)\left(|10\rangle + |01\rangle\right)\Bigr] \\
    &= \frac{1}{2}\Bigl[ |00\rangle\left(\alpha|0\rangle + \beta|1\rangle\right) + |01\rangle\left(\alpha|1\rangle + \beta|0\rangle\right) \\
    &~~~+ |10\rangle\left(\alpha|0\rangle - \beta|1\rangle\right) + |11\rangle\left(\alpha|1\rangle - \beta|0\rangle\right) \Bigr].
    \end{aligned}
\end{equation}

\emph{Dynamic quantum circuits} refer to quantum circuits with mid-circuit measurements and feedforward operations~\cite{dqc_corcoles2021exploiting}.
Recent experimental studies have demonstrated the potential of DQCs in various application scenarios, including short-depth state preparation~\cite{smith2024constant_depth_mps_dqc}, device scale expansion~\cite{vazquez2024scaling_qc_with_dqc}, and quantum fan-out gate implementation~\cite{baumer2024dqc_qft}.

The power of DQC can be illustrated by \emph{quantum teleportation}, a technique for moving quantum states~\cite{nielsen2010qc_and_qi}.
Fig.~\ref{fig:quantum_tele} shows the circuit diagram for teleporting a qubit from Alice's system (the top two lines) to Bob's system (the bottom line).
Initially, Alice and Bob together generated an EPR pair and each of them takes one qubit, that is, $|\beta_{00}\rangle = \frac{1}{\sqrt{2}}\left( |00\rangle + |11\rangle \right)$.
Then the mission of Alice is to teleport an unknown state $|\psi\rangle = \alpha|0\rangle + \beta|1\rangle$ to Bob by sending only classical information, which can be achieved by the gates and measurement feedforward operations.
Specifically, Alice applies a CNOT gate and a Hadamard gate to her qubits and send the measurement results of $|\psi_2\rangle$ -- as derived and shown on the right side of Fig.~\ref{fig:quantum_tele} -- to Bob.
Depending on Alice's measurement outcome, Bob could recover the state $|\psi\rangle$ by applying corresponding operations.
For example, if the measurement result is 01 then Bob can fix up his state by applying the $X$ gate.

\subsection{Quantum Control Systems}

\label{ssec:qcs}

\yilun{describe the execution of a DQC on QCPs}

Although the immense value of DQCs has long been recognized, only recent advancements in control systems have enabled DQCs to be flexibly programmed and executed on real machines, given their stringent requirements for real-time classical processing capabilities~\cite{cross2022openqasm3,IBM_dqc}.
A \emph{quantum control system} is a specialized classical system composed of hardware and software designed to control quantum devices.
Quantum circuits are compiled into quantum control instructions and then executed by the control system~\cite{fu2019eqasm}.
Therefore, it is easy to comprehend the following insight:

\begin{insightbox}{Insight}
    Quantum program execution time is determined by the instruction processing time in the quantum control system.
\end{insightbox}

\begin{figure}[h]
\centering
    \includegraphics[width=\linewidth]{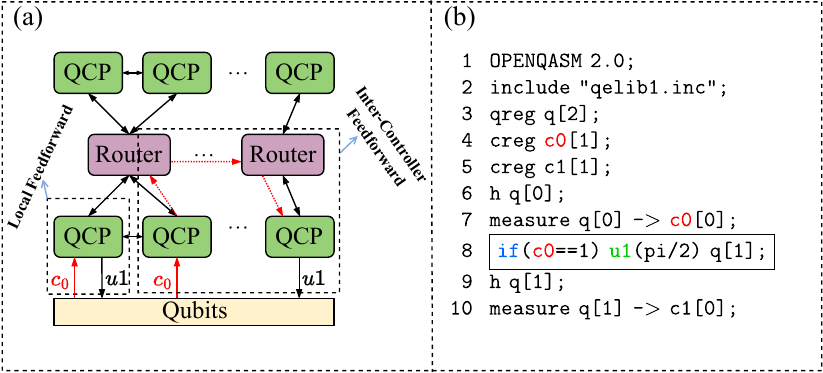}
\caption{(a) Schematic of a distributed quantum control system. (b) Source program of dynamic quantum Fourier transform~\cite{baumer2024dqc_qft,chen2022veriqbench}.}
\label{fig:dist_qcs}
\end{figure}

With the number of qubits increases, quantum control systems evolve to a distributed architecture as shown in Fig.~\ref{fig:dist_qcs}(a), which includes a group of QCPs interconnected via some routers~\cite{ibm_sys_isscc_zettles202226,ali_cl_arch_zhang2023classical}.

Taking the program shown in Fig.~\ref{fig:dist_qcs}(b) as an example, performing a measurement feedforward task in the control system can involve diverse communication paths
determined by the mapping between qubits and controllers.
For example, the local feedforward block represents a scenario where $q_0$ and $q_1$ are managed by the same QCP.
In this case, the feedforward process involves no inter-controller communication.
By contrast, if qubits $q_0$ and $q_1$ are managed by separate QCPs (the inter-controller feedforward block), the measurement result $C_0$ may involve multiple communication hops.

%% file: paper/2-motivation.tex
\subsection{Problem and Motivation}

\label{sec:problem_and_motivation}

\begin{figure}[h!]
    \centering
        \includegraphics[width=\linewidth]{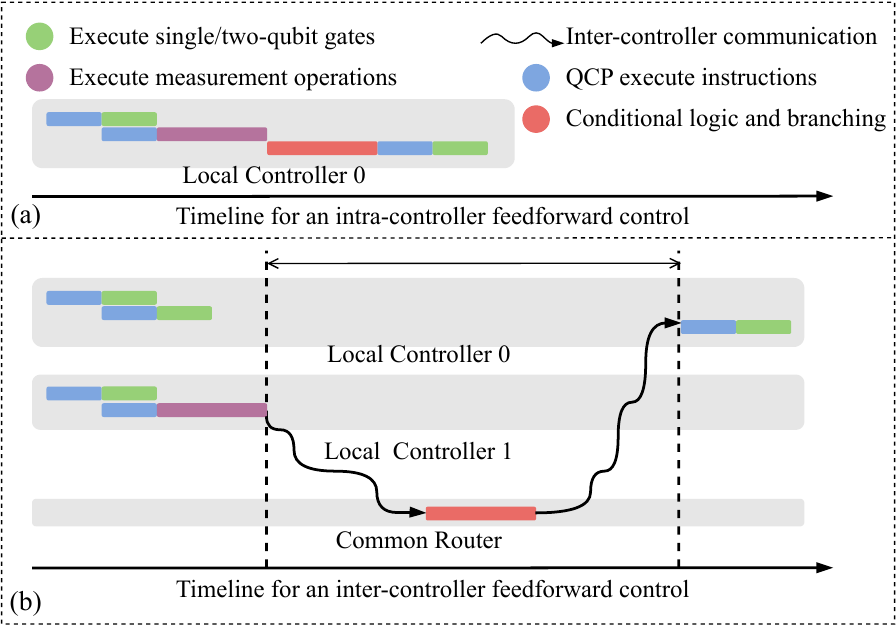}
    \vspace{-5pt}
    \caption{The latency breakdown for intra- and inter-controller feedforward.}
    \vspace{-10pt}
\label{fig:motivation_pipeline}
\end{figure}

Current quantum devices are constrained by limited qubit lifetimes, typically on the order of hundreds of microseconds for state-of-the-art superconducting qubits.
Reducing the execution time of quantum programs to mitigate errors from decoherence has thus become a key objective.
The lifecycle of a quantum control instruction can be divided into two stages based on current QCP designs~\cite{fu2017quma,fu2019eqasm}: (i) fetching, decoding, and queuing the quantum event until it is emitted to the quantum-classical interface (QCI), represented by blue blocks;
and (ii) executing pulse waveforms on quantum devices, where quantum gates are shown as green blocks and measurements, including acquisition and waiting for results, are represented by purple blocks.
For static circuits without mid-circuit measurements, the execution time of quantum programs can be effectively approximated by the quantum device time, as the processing time on QCPs is pipelined with the pulse durations on the quantum device.
Traditional LSQC techniques thus focus on minimizing circuit depth and SWAP overhead to reduce quantum device time.
However, this approach is insufficient for DQCs, where a quantum gate instruction may depend on prior measurement results and associated classical processing and QCP time and thus cannot be well-hidden by pulse durations on the quantum device (Fig.~\ref{fig:motivation_pipeline}(a)).
Therefore, it is necessary to reshape the existing LSQC design methodology from device-centric to controller-centric.

In this work, we identify a unique optimization opportunity for the layout synthesis of DQCs.
Specifically, intra-controller feedforward, where measurement and dependent qubits are managed locally by the same controller, avoids inter-controller communication, reducing latency (Fig.~\ref{fig:motivation_pipeline}(a)).
In contrast, inter-controller feedforward potentially requires communications between different controllers with extended latency (Fig.~\ref{fig:motivation_pipeline}(b)).
Existing studies validate this latency difference.
For example, Ref.~\cite{ibm_ctrl_sys_nature_gupta2024encoding} reports a branching latency of $\sim$500 ns in distributed systems with a central router, while intra-controller feedforward latency is as low as 92 ns~\cite{fu2019eqasm} and can reach 50 ns in leading industry products~\cite{zurich_instrument}.
Motivated by this discrepancy, this work aims to design a layout synthesizer that minimizes inter-controller communication latency.

%% file: paper/3-design-0.tex
\section{Approach}

In this section, we present the design of \name{}.
We begin with \emph{Type-I DQCs}, which are characterized by the absence of connectivity constraints and the replacement of CNOT gates with measurement and feedforward operations, significantly reducing circuit depth.
A notable example is the dynamic circuit for quantum Fourier transform (QFT)~\cite{griffiths1996semiclassical_dqc_qft},
a core subroutine in numerous quantum algorithms~\cite{shor1999polynomial,paler2022quantum_qft_addition}.
Next, we address \emph{Type-II DQCs}, which are subject to connectivity constraints.
For these circuits, the LSQC objectives are to minimize both inter-controller communication latency and circuit depth.

\subsection{Type-I DQC}

\label{ssec:type_1_dqc}

\begin{figure}[h!]
\centering
\includegraphics[width=\linewidth]{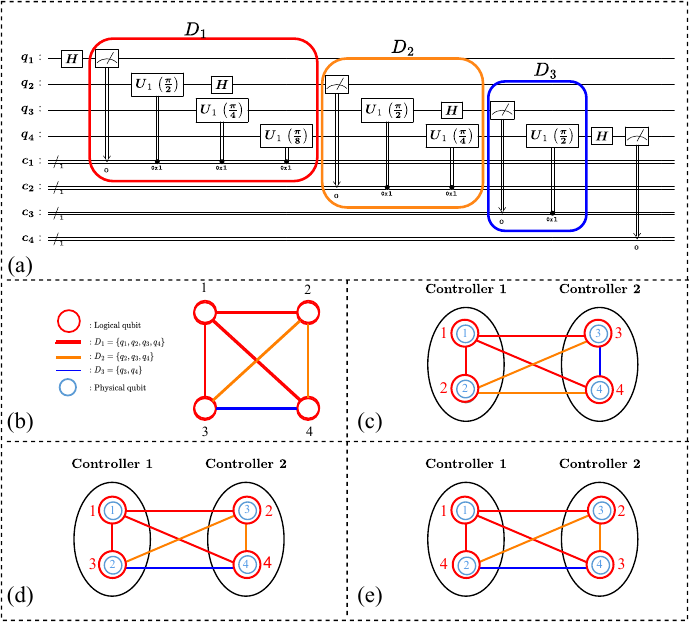}
\caption{An illustrational example of the minimum-cut graph-partitioning problem based on the 4-qubit DQC of QFT. (a) Circuit diagram of a 4-qubit dynamic QFT; (b) Graph representation of all feedforward operations; (c)-(e) Three different logical-to-physical qubit mapping schemes: $\mathcal{M}_1^Q,\mathcal{M}_2^Q,\mathcal{M}_3^Q$.}
\label{fig:graph_cut_walk_through}
\end{figure}

\subsubsection{Motivational Example}

To better understand the modeling approach of \name{}, we first describe an illustrational example based on a DQC-implementation of QFT as shown in Fig.~\ref{fig:graph_cut_walk_through}(a).
We denote logical qubits as \( q_1, q_2, \dots, q_n \), where \( n \) is the total number of qubits in the circuit.

\squishlist{}

\item \textbf{Graphical representation of feedforward operations.}
In Fig.~\ref{fig:graph_cut_walk_through}(b), the feedforward operations from the example circuit are extracted and represented as a graph.
Nodes correspond to logical qubits, and an edge between two qubits indicates that an operation on one depends on the measurement of the other.
For example, the gates \( \mathrm{U}_1\left(\frac{\pi}{2}\right), \mathrm{U}_1\left(\frac{\pi}{4}\right), \) and \( \mathrm{U}_1\left(\frac{\pi}{8}\right) \), acting on \( q_2, q_3, \) and \( q_4 \), are all conditioned on the measurement of \( q_1 \), resulting in three red edges: $\{q_1,q_2\}, \{q_1,q_3\}, \{q_1,q_4\}$.
The involved qubits form a \emph{conditional inter-dependent qubits} (CIDQ) set, denoted as \( D_1 = \{q_1, q_2, q_3, q_4\} \).
Two other CIDQ sets can be similarly identified: \( D_2 = \{q_2, q_3, q_4\} \) and \( D_3 = \{q_3, q_4\} \).
In a CIDQ set \( D_i \), the measured qubits form \( D_i^m \), and the target qubits that depend on those measurements form \( D_i^t \).
For instance, in \( D_1 \), we have \( D_1^m = \{q_1\} \) and \( D_1^t = \{q_2, q_3, q_4\} \).

\item \textbf{Logical-to-physical qubit mapping schemes.}
Fig.~\ref{fig:graph_cut_walk_through}(c)-(e) depict different logical-to-physical qubit mapping schemes.
Consider two controllers $C_1$ and $C_2$ that manage physical qubits $\{Q_1,Q_2\}$ and $\{Q_3,Q_4\}$ respectively.
There is no distinction between (i) \(\{q_1, q_2\} \to C_1, \{q_3, q_4\} \to C_2\) and (ii) \(\{q_1, q_2\} \to C_2, \{q_3, q_4\} \to C_1\).
Consequently, a little thought shows that there are only three possible logical-to-physical mappings:

\begin{equation}
\footnotesize
\begin{aligned}
\text{(i)} & \quad \mathcal{M}^Q_1 \equiv \{q_1, q_2\} \to \{Q_1, Q_2\}, \{q_3, q_4\} \to \{Q_3, Q_4\}; \\
\text{(ii)} & \quad \mathcal{M}^Q_2 \equiv \{q_1, q_3\} \to \{Q_1, Q_2\}, \{q_2, q_4\} \to \{Q_3, Q_4\}; \\
\text{(iii)} & \quad \mathcal{M}^Q_3 \equiv \{q_1, q_4\} \to \{Q_1, Q_2\}, \{q_2, q_3\} \to \{Q_3, Q_4\}.
\end{aligned}
\end{equation}

\squishend{}

Our goal is to find a logical-to-physical mapping that minimizes inter-controller communication.
Under $\mathcal{M}_1^Q$, the first ($D_1$) and second ($D_2$) feedforward operations require sending measurement results of $q_1$ and $q_2$ from $C_1$ to $C_2$, while the third ($D_3$) is performed locally on $C_2$.
Thus, $\mathcal{M}_1^Q$ incurs 2 communication steps between $C_1$ and $C_2$.
In comparison, $\mathcal{M}_2^Q$ and $\mathcal{M}_3^Q$ each result in 3 communication steps.
Therefore, $\mathcal{M}_1^Q$ is the optimal mapping with the lowest communication overhead.
Interestingly, as shown in Fig.~\ref{fig:graph_cut_walk_through}(c)-(e), the number of distinct edge colors crossing the two controllers matches the number of communication steps.
This observation forms the basis of our approach: transforming the LSQC problem into an equivalent graph-cut problem:

\begin{insightbox}{Core Idea}
    Layout synthesis for minimizing inter-controller communication latency is equivalent to a minimum-cut graph partitioning problem.
\end{insightbox}

\subsubsection{Problem Formulation}

The minimum-cut graph partitioning problem can be formally described as follows.

\squishlist{}

\item \textbf{Hypergraph definition.}
Given a DQC, each feedforward operation is extracted as a CIDQ set, forming a list \( L^D \), where \( L^D[i,j] \) denotes the $j$th qubit in the $i$th CIDQ set \( D_i \).
Using \( L^D \), we construct a hypergraph \( U(V,E) \) representing all feedforward dependencies, as illustrated in Fig.~\ref{fig:graph_cut_walk_through}(b).
Each node in \( V \) corresponds to a logical qubit, and each hyperedge in \( E \) represents a CIDQ set connecting multiple nodes.

\item \textbf{Graph partitioning.}
Consider a distributed quantum control system with $k$ controllers \( \{C_i\}_{i=1}^k \) managing $m$ physical qubits \( \{Q_i\}_{i=1}^m \).
Each controller is responsible for a subset of physical qubits, defined by a mapping function \(\mathcal{M}^C \equiv f: \{Q_i\} \to \{C_j\}\).
The hypergraph \( U \) is partitioned into subgraphs, each assigned to a distinct controller.
As a result, some hyperedges are cut across multiple controllers.
Note that, a specific partitioning scheme is determined by the logical-to-physical qubit mapping $\mathcal{M}^Q$, which is the output of our layout synthesizer.

\item \textbf{Optimization objective.}
Cut hyperedges correspond to inter-controller communication.
Our objective is finding a mapping $\mathcal{M}^Q$ to minimize the total communication latency, defined as the sum of latencies associated with all cut edges.

\begin{equation}
\mathcal{L}(\mathcal{M}^Q) \equiv \sum_{D_i \in L^D} S(D_i).
\end{equation}

\noindent
Here, \( S(D_i) \) denotes the communication latency for sending measurement results from \( D_i^m \) to \( D_i^t \), which depends on the topology and communication protocol of the control system.
In our implementation, we use the minimum number of communication hops---determined by the controller topology---to represent this latency, which we refer to as \emph{Inter-Controller Communication Steps} (ICCS).

\squishend{}

\begin{figure*}[h!]
    \centering
    \includegraphics[width=\linewidth]{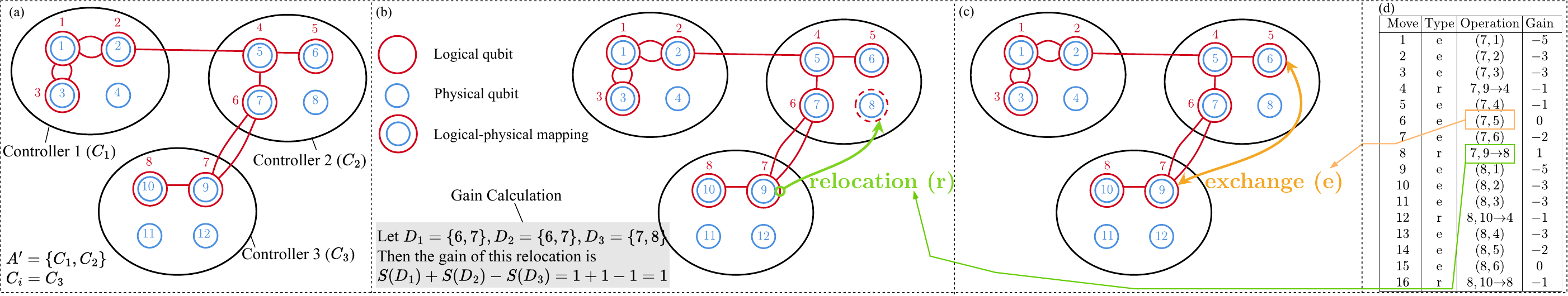}
    \caption{Examples of different movements and gain calculation during a qubit moving pass (line 7 in Alg.~\ref{alg:qubit_move_pass}). For illustration purpose, we assume all CIDQ sets involve only two qubits and the ICCS between any pair of controllers is uniformly 1. (a) A possible logical-to-physical mapping $\mathcal{M}^Q_{current}$. (b) Example of qubit relocation and its gain calculation process.
    After relocating $q_7$ to $Q_8$, $D_1$ and $D_2$ belong to the same controller, while $D_3$ is cut by two controllers.
    Therefore, the gain is calculated by $1+1-1 = 1$.
    (c) Example of qubit exchange. (d) Table of gains of all movements between $C_3$ and $A^\prime = \{C_1,C_2\}$.}
    \label{fig:qubit_moving_pass}
\end{figure*}

\begin{table}
\centering
\caption{List of symbols used in our approach.}
\label{tab:symbols}
\begin{tabular}{|>{\centering\arraybackslash}m{1.6cm}|m{6cm}|}
\hline
\textbf{Symbol} & \centering\arraybackslash\textbf{Description} \\
\hline
$D_i$ & A specific CIDQ set representing a group of conditionally interdependent qubits. \\
\hline
 $D_i^m$ & Subset of qubits in $D_i$ that are measured. \\
\hline
 $D_i^t$ & Subset of qubits in $D_i$ that require measurements from other qubits. \\
\hline
 $\mathcal{M}^Q$ & Logical-to-Physical qubit mapping function: Maps logical qubits to physical qubits on a device. \\
\hline
 $\mathcal{M}^C$ & Qubit-Controller mapping function: Maps physical qubits to their controlling QCPs. \\
\hline
 $L_D$ & A list of CIDQ sets derived from a given DQC. \\
\hline
 $S(D_i)$ & Communication cost for executing feedforward operations in $D_i$. \\
\hline
 $\mathcal{L}(\mathcal{M}^Q)$ & Objective function to minimize: The sum of $S(D_i)$ over all CIDQ sets. \\
\hline
 $U$ & Undirected hypergraph where vertices represent logical qubits and hyperedges represent CIDQ sets, used to transform the LSQC problem into a graph partitioning task.\\
\hline
\end{tabular}
\end{table}

\subsubsection{Algorithm Design}

While the example in Fig.~\ref{fig:graph_cut_walk_through} partitions graph vertices into equally sized subsets, our problem does not require such balance.
For instance, in a DQC with 12 qubits and two controllers, each managing up to 10 qubits, assigning the two qubits involved in the fewest CIDQ sets to one controller and the remaining 10 qubits to the other may yield better results than an even split.
This differentiates our problem from traditional \(k\)-way graph-partitioning problems~\cite{lee2024gkway}.
To address this, we propose a two-stage heuristic approach (Alg.~\ref{alg:iterative_kl_mapper}), detailed as follows.

\begin{figure}[t]
    \centering
    \includegraphics[width=0.4\linewidth]{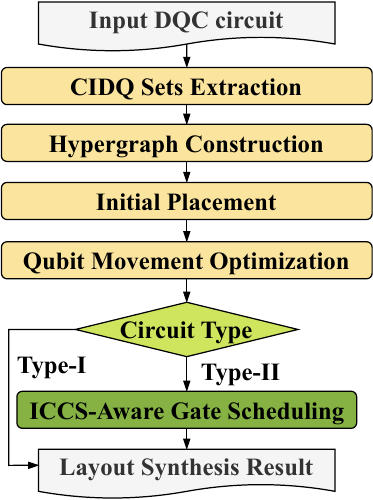}
    \caption{The layout synthesis flow of \name{}.}
    \vspace{-4pt}
    \label{fig:flow}
\end{figure}

\begin{algorithm}[h!]
    \scriptsize
    \SetAlgoLined
    \SetKwInOut{KwIn}{Input}
    \SetKwInOut{KwOut}{Output}
    \SetKwFunction{GenerateInitialMapping}{gen\_initial\_mapping}
    \SetKwFunction{sort}{sort}
    \SetKwFunction{update}{update}
    \SetKwFunction{ConstructGraph}{construct\_graph}
    \SetKwFunction{KlPass}{apply\_qubit\_moving\_pass}

    \KwIn{Controller-Qubit Mapping $\mathcal{M}^C$, List of CIDQ Sets $L^D$}
    \KwOut{Logical-Physical Mapping $\mathcal{M}^Q$}

    \tcp{\textcolor{blue}{Stage 1: }Initialize $\mathcal{M}^Q$}

    $U \gets$\ConstructGraph{$L^D$}\;
    Initialize $\mathcal{M}^Q(q_i)$ as -1 for all qubits\;
    \ForEach{logical qubit $q_i$ in the descending order of degrees in $U$}{
        $controller\_score \gets \{\}$\;
        \ForEach{$q_j$ in the neighbors of $q_i$ in $U$}{
            $Q_j \gets \mathcal{M}^Q(q_j)$\;
            \If{$Q_j \neq -1$}{
                $C_j \gets \mathcal{M}^C(Q_j)$\;
                $controller\_score[C_j]\gets controller\_score[C_j] + 1$\;
            }
        }
        \eIf{$controller\_score$ is not empty}{
            Find $C_i$ with the maximum score in $controller\_score$\;
        }{
            Randomly choose a controller $C_i$ for $q_i$ and its neighbors\;
        }
        Randomly choose a physical qubit $Q_i$ from the physical qubits obtained from the inverse mapping of $\mathcal{M}^C$\;
        $\mathcal{M}^Q$.\update{$q_i \to Q_i$}\;
    }

    \tcp{\textcolor{blue}{Stage 2:} Iteratively move qubits across the allocated controllers to further reduce the number of ICCSs}

    $best\_score \gets \infty$\;
    \ForEach{controller $C_i$ in all $k$ controllers $A\equiv \{C_a\}_{a=1}^k$}{
        $A^\prime \gets A \backslash  \{C_i\}$; \tcp{Obtain the controllers excluding $C_i$}
        $\mathcal{M}^Q_{temp}\gets$ \KlPass{$\mathcal{M}^Q,C_i,A^\prime$}\;
        $current\_score \gets \mathcal{L}(\mathcal{M}^Q_{temp})$\;
        \If{$current\_score < best\_score$}{
            $\mathcal{M}^Q\gets \mathcal{M}^Q_{temp}$\;
            $best\_score \gets current\_score$ \;
        }
    }
    \caption{ICCS-Aware Initial Placement}
    \label{alg:iterative_kl_mapper}
\end{algorithm}

\setlength{\textfloatsep}{8pt}
\begin{algorithm}[h!]
    \scriptsize
    \SetAlgoLined
    \SetKwInOut{KwIn}{Input}
    \SetKwInOut{KwOut}{Output}
    \SetKwFunction{GenerateInitialMapping}{gen\_initial\_mapping}
    \SetKwFunction{ObtainAllMovements}{obtain\_movements}
    \SetKwFunction{update}{update}
    \SetKwFunction{append}{append}
    \SetKwFunction{remove}{remove}
    \SetKwFunction{ConstructGraph}{construct\_graph}
    \SetKwFunction{KlPass}{apply\_qubit\_moving\_pass}

    \KwIn{Logical-Physical Mapping $\mathcal{M}^Q$, Controller $C_i$, Set of Other Controllers $A^\prime$}
    \KwOut{New Logical-Physical Mapping $\mathcal{M}^Q_{temp}$}

    $\mathcal{M}^Q_{temp} \gets \mathcal{M}^Q$\;
    $\mathcal{M}^Q_{current} \gets \mathcal{M}^Q$\;
    $best\_movements \gets []$\;
    $gains \gets []$\;
    $movements \gets $\ObtainAllMovements{$C_i,A^\prime$}\;
    \While{$movements \neq \emptyset$}{
        Find a $move$ that has the maximal gain $g$ based on $\mathcal{M}^Q_{current}$\;
        $movements$.\remove{$move$}\;
        $best\_movements$.\append{$move$}\;
        $gains$.\append{$g$}\;
        $\mathcal{M}^Q_{current}.\update{move}$\;
    }
    Find a $l$ that maximize $max\_gain \gets \sum_{j=1}^l gains[i]$\;
    \If{$max\_gain > 0$}{
        $\mathcal{M}^Q_{temp}$.\update{$best\_movements[1:l]$}\;
    }

    \caption{Qubit Moving Pass}
    \label{alg:qubit_move_pass}
\end{algorithm}

\squishlist{}
\item \textbf{Stage 1: Initialize $\mathcal{M}^Q$ via greedy allocation of controllers.}
First, we construct an undirected graph \( U \) based on \( L^D \), where the vertices in \( U \) correspond to logical qubits in a DQC.
The graph \( U \) is a \emph{hypergraph}, allowing edges to connect multiple vertices, with each CIDQ set \( D_i \) represented as a hyperedge in \( U \).
Next, we traverse the vertices in \( U \) in descending order of their \emph{degrees} -- defined as the number of CIDQ sets in which a qubit is included -- to \emph{prioritize the allocation of highly interdependent qubits to the same controller.}
For each vertex (qubit), if its neighbors have not been assigned to any controller, we randomly map this qubit and its neighbors to the physical qubits of a controller.
Otherwise, we identify the controllers managing its neighboring qubits and map it to the controller responsible for the most neighbors (lines 2-14).

\item \textbf{Stage 2: Iteratively move qubits across the allocated controllers to further reduce the number of ICCSs.}
Let \( A \equiv \{C_a\}_{a=1}^k \) denote the set of all \( k \) controllers.
For each controller \( C_i \) in \( A \), we exclude \( C_i \) to form a new set \( A^\prime \) (line 18) and perform a \emph{qubit moving pass} between \( C_i \) and \( A^\prime \), generating a temporary mapping \(\mathcal{M}^Q_{temp}\) (line 19).
The mapping with the lowest objective function value \(\mathcal{L}(\mathcal{M}^Q_{temp})\) is selected as the final mapping (lines 20–23).
Inspired by the classic Kernighan–Lin (KL) algorithm~\cite{kernighan1970KL_partition}, which searches between two partitions, our approach (Alg.~\ref{alg:qubit_move_pass}) generalizes this by exploring movements between one controller (\( C_i \)) and all remaining controllers (\( A^\prime \)), thereby significantly expanding the search space.
Furthermore, unlike KL’s bidirectional exchanges, our \emph{hybrid movement design} includes both bidirectional exchanges of qubits between \( C_i \) and \( A^\prime \) (Fig.~\ref{fig:qubit_moving_pass}(c)), as well as unidirectional relocations of a single qubit from \( C_i \) to another controller in \( A^\prime \), provided this does not exceed the capacity constraints of the target controller (Fig.~\ref{fig:qubit_moving_pass}(b)).
This design is enabled by the absence of a balance constraint in our problem, thus the relocation of a single qubit allows our algorithm to explore solution spaces of unbalanced partitions.
The \emph{gain} of a movement, defined as the change in the sum of ICCSs of the affected CIDQ sets, is shown in Fig.~\ref{fig:qubit_moving_pass}(d), based on the initial mapping in Fig.~\ref{fig:qubit_moving_pass}(a).
In this example, a unidirectional relocation achieves the highest gain, demonstrating the effectiveness of incorporating hybrid movements and highlighting the potential benefits of removing balance constraints.
In each iteration, we iteratively select, apply, and remove the movement with the greatest gain until all are considered (lines 6–11).
We then update \(\mathcal{M}^Q_{temp}\) with the maximum accumulated gain and finalize the mapping (lines 12–14).

\squishend{}

\subsubsection{Complexity Analysis}

In Stage 1, for each qubit (with \( n \) in total), we traverse its neighboring qubits (with \( d \) on average) and identify the controllers allocated to these qubits.
Then, we determine the controller that manages the largest number of neighboring qubits among all \( k \) controllers.
Thus, the complexity of Stage 1 is \( O(n(k + d)) \).
In Stage 2, for each controller (with \( k \) in total), we perform a qubit moving pass, where the complexity is determined by the number of movements.
The average number of qubits managed by a controller is \(\frac{n}{k}\).
Since each qubit can either be moved from \( C_i \) to any of the other controllers in \( A^\prime \) or exchanged with any other qubit in \( A^\prime \), the number of movements per qubit is \( (n + k) \).
Therefore, the total number of movements per pass is \( N = \frac{n}{k}(n + k) \).
These movements can be stored in a priority queue, with the movement of maximum gain (\( move \)) placed at the front.
After applying \( move \), the gains of movements associated with the qubits involved in \( move \) need to be recalculated.
The total number of such associated movements is \( M = d(n + k) \).
Since updating an element in a priority queue has a complexity of \( O(\log(N)) \), the complexity of updating gains is \( O(M \cdot \log(N)) \).
Finally, the total complexity of stage 2 is \( O(k \cdot N \cdot M \cdot \log(N)) = O(k \cdot \frac{n}{k}(n + k) \cdot d(n + k) \cdot \log(\frac{n}{k}(n + k))) =  O(dn(n+k)^2log(\frac{n}{k}(n+k)))\).

%% file: paper/3-design-1.tex
\begin{algorithm}[b!]
    \scriptsize
    \SetAlgoLined
    \SetKwInOut{KwIn}{Input}
    \SetKwInOut{KwOut}{Output}
    \SetKwFunction{GetMinScoreSwaps}{obtain\_min\_score\_swaps}
    \SetKwFunction{GetSwaps}{obtain\_swaps}
    \SetKwFunction{len}{len}
    \SetKwFunction{update}{update}
    \SetKwFunction{GetCidqSets}{obtain\_cidq\_sets}
    \SetKwFunction{CostDepth}{obtain\_depth\_cost}
    \SetKwFunction{RandChoice}{random\_choice}

    \KwIn{Controller-Qubit Mapping $\mathcal{M}^C$, List of CIDQ Sets $L^D$, Initial Logical-Physical Mapping $\mathcal{M}^Q$, Front Layer $F$, Device Coupling Map $G(V,E)$}
    \KwOut{Inserted SWAPs, Final Logical-Physical Mapping $\mathcal{M}^Q_f$}

    \While{$F$ is not empty}{
        $execute\_gate\_list \gets \emptyset$\;
        Find executable gates in $F$ and put them into $execute\_gate\_list$\;
        \eIf{$execute\_gate\_list \neq \emptyset$}{
            \ForEach{gate in $execute\_gate\_list$}{
                Remove gate from $F$ and put its successors in DAG into $F$ if the gates' dependencies are resolved\;
            }
            continue;
        }{
            $score = \{\}$\;
            $swap\_candidate\_list$ = \GetSwaps{$F,G$}\;
            \For{$swap$ in $swap\_candidate\_list$}{
                $\mathcal{M}^Q_{temp} = \mathcal{M}^Q$.\update{$swap$}\;
                $score[swap] \gets$ \CostDepth{$F, \text{DAG}, G, \mathcal{M}^Q_{temp}$}\;
            }
            $similar\_swaps \gets $\GetMinScoreSwaps{$score$}\;
            \tcp{\textcolor{blue}{ICCS-Aware Search}}
            \eIf{\len{$similar\_swaps$} $> 1$}{
                $iccs\_score \gets \{\}$\;
                \For{$swap$ in $similar\_swaps$}{
                    \tcp{Calculate ICCS score for these SWAPs}
                    $\mathcal{M}^Q_{temp} \gets$ $\mathcal{M}^Q$.\update{$swap$}\;
                    $L^D_{active} \gets$ \GetCidqSets{$F,DAG,G,\mathcal{M}^Q_{temp}$}\;
                    $iccs\_score[swap] \gets \sum_{D_i\in L^D_{active}}S(D_i)$\;
                }
                $swaps \gets$ \GetMinScoreSwaps{$iccs\_score$}\;
                $swap \gets$ \RandChoice{$swaps$}\;
            }{
                $swap \gets similar\_swaps[0]$\;
            }
            $\mathcal{M}^Q$.\update{$swap$}\;
        }
    }
    \caption{ICCS-Aware Gate Scheduling}
    \label{alg:iccs_sabre_search}
\end{algorithm}

\subsection{Type-II DQCs}

Connectivity constraints add complexity to minimizing inter-controller communication.
Since two-qubit gates require physically adjacent qubits, SWAP gates are often inserted to enable gate execution—a process known as \emph{gate scheduling}~\cite{tan2020OLSQ} or \emph{qubit routing}~\cite{fu2023effective_and_efficient_mapper}.
Prior work has largely focused on reducing circuit depth or the number of SWAPs, without considering the underlying controller architecture.
However, inserting SWAPs may inadvertently split a CIDQ set across multiple controllers, increasing \# ICCS.

To address this challenge, we design a latency-aware gate scheduler that minimizes both the number of additional SWAP gates and inter-controller latency.
The core idea is to \emph{use inter-controller latency as a tie-breaker when multiple SWAP insertion options have the same cost.}
Existing heuristic layout synthesizers, such as those in Refs.~\cite{li2019tackling,zhang2021timeoptimal_qubit_mapping,fu2023effective_and_efficient_mapper}, typically use greedy search strategies that evaluate potential SWAP costs at each step by considering future SWAP insertion possibilities based on predefined cost functions.
In some cases, multiple SWAP options may yield identical or nearly identical costs.
This creates an opportunity to incorporate a controller-latency metric to proactively guide SWAP selection, avoiding random choices.
Notably, this design philosophy can be generalized to all existing gate schedulers, enabling them to account for controller communication latency during the decision-making process of whether and where to insert SWAPs.
As a demonstration, we integrate our strategy into the gate scheduling stage of SABRE~\cite{li2019tackling} as it is widely used in the community.
Alg.~\ref{alg:iccs_sabre_search} describes the process of ICCS-aware SABRE search,
where the directed acyclic graph (DAG) represents the execution dependencies of operations in the circuit.
The front layer \( F \) is defined as the set of all two-qubit gates with no unexecuted predecessors in the DAG,
and the coupling map \( G \) represents the topology of the target quantum device.
For a comprehensive explanation of its modeling methodology, readers may refer to the original SABRE paper~\cite{li2019tackling}.
Here, we omit some details of the original SABRE steps and focus on explaining the relevant modifications that make it aware of ICCS.
When there are no executable gates in \( F \), we first identify all possible SWAPs associated with the qubits in \( F \) and calculate a score for each SWAP to estimate its negative impact on circuit depth (lines 9$\sim$13).
Next, we initiate an ICCS-aware search procedure when multiple SWAPs yield identical scores (lines 15–22).
For each SWAP, we first obtain a temporary mapping \(\mathcal{M}^Q_{temp}\) by applying the SWAP.
We then look ahead to identify a set of feedforward operations whose dependencies in the DAG are resolved after applying the SWAP and extract the corresponding CIDQ sets from \( L^D \) to construct a new list, \( L^D_{active} \).
Subsequently, we calculate the sum of the ICCSs associated with each CIDQ set in \( L^D_{active} \) to determine the \( iccs\_score \) of the SWAP.
The SWAP with the lowest \( iccs\_score \) is selected as the final choice.

%% file: paper/4-evaluation.tex
\section{Evaluation\yilun{}}

\label{sec:eval}

\subsection{Experiment Setup}

\subsubsection{Implementation}

We implement \name{} as a framework that interfaces with Qiskit. \name{} consists of $\sim$3k lines of Python code and around $\sim$1k lines of modifications in the Rust library of Qiskit.
For Type-I DQCs, we set the outcome of our initial placement as the initial layout of Qiskit transpiler.
For Type-II DQCS, we extend the SABRE implementation in Qiskit and use the initial placement as the starting layout.

\subsubsection{Benchmarks}
Benchmarks are collected from VeryQBench~\cite{chen2022veriqbench} and QASMBench~\cite{li2023qasmbench}, including dynamic QFT, iterative phase estimation (PE), and the quantum counterfeit coin (CC) problem.
Additionally, we construct randomized DQCs to cover a broader range of circuit patterns by using the blocks from the randomized benchmarking protocol~\cite{shirizly2024rb_dqc} as basic components (Random).

\subsubsection{Metrics}
Several post-compilation metrics are collected for performance comparison with Qiskit as our baseline, including circuit depth, number of operations, and number of ICCSs (\# ICCS).

\subsubsection{System Configurations}
In our main results, we adopt a star-topology controller architecture, where all controllers are connected to a central router~\cite{ibm_sys_isscc_zettles202226}.
This topology is chosen as, to the best of our knowledge, it is the only publicly available solution that provides both a concrete controller topology design and support for arbitrary feedforward operations.
The qubit device architecture is based on IBM's 127-qubit quantum processor, which features a heavy-hex-lattice topology.
All experiments were performed on a Linux server with 768~GB of memory and two 32-core Intel(R) Xeon(R) Silver 4216 CPUs.

\begin{table}[h!]
    \centering
    \caption{Comparison between \name{} and baseline ($k=4$).}
    \scalebox{0.7}{
    \begin{tabular}{llcccccc}
        \toprule
        \textbf{Benchmark} & \textbf{Qubits} & \multicolumn{3}{c}{Baseline} & \multicolumn{3}{c}{\name{}} \\
        \cmidrule(lr){3-5} \cmidrule(lr){6-8}
        & & \# Operations & Depth & \# ICCS & \# Operations & Depth & \# ICCS \\
        \midrule
        \multicolumn{8}{l}{\textit{Type-I Benchmarks}} \\
        qft & 20 & 270 & 99 & 144 & 270 & 99 & 0 \\
        qft & 30 & 555 & 149 & 335 & 555 & 149 & 0 \\
        qft & 40 & 940 & 199 & 599 & 940 & 199 & 256 \\
        qft & 50 & 1425 & 249 & 933 & 1425 & 249 & 576 \\
        \midrule
        \multicolumn{8}{l}{\textit{Type-II Benchmarks}} \\
        cc & 12 & 159 & 109 & 24 & 153 & 110 & 6 \\
        cc & 32 & 487 & 315 & 120 & 586 & 346 & 6 \\
        pe & 20 & 441 & 171 & 125 & 434 & 184 & 19 \\
        pe & 30 & 798 & 268 & 309 & 870 & 277 & 29 \\
        pe & 40 & 1356 & 400 & 591 & 1444 & 430 & 309 \\
        pe & 50 & 1938 & 505 & 911 & 2101 & 505 & 619 \\
        random & 20 & 2096 & 726 & 339 & 2095 & 785 & 39 \\
        random & 30 & 5551 & 1589 & 882 & 5738 & 1852 & 111 \\
        random & 40 & 11480 & 2810 & 1500 & 11413 & 3016 & 912 \\
        random & 50 & 19743 & 4710 & 2691 & 20140 & 4621 & 1812 \\
        \midrule
        \textit{Type-II Average} & - & 4404.90 & 1160.30 & 749.20 & 4497.40 & 1212.60 & 386.20 \\
        \bottomrule
    \end{tabular}
    }
    \label{tab:main_res}
\end{table}

\subsection{Performance on Type-I DQCs}

\begin{figure}[ht]
    \centering
    \includegraphics[width=0.7\linewidth]{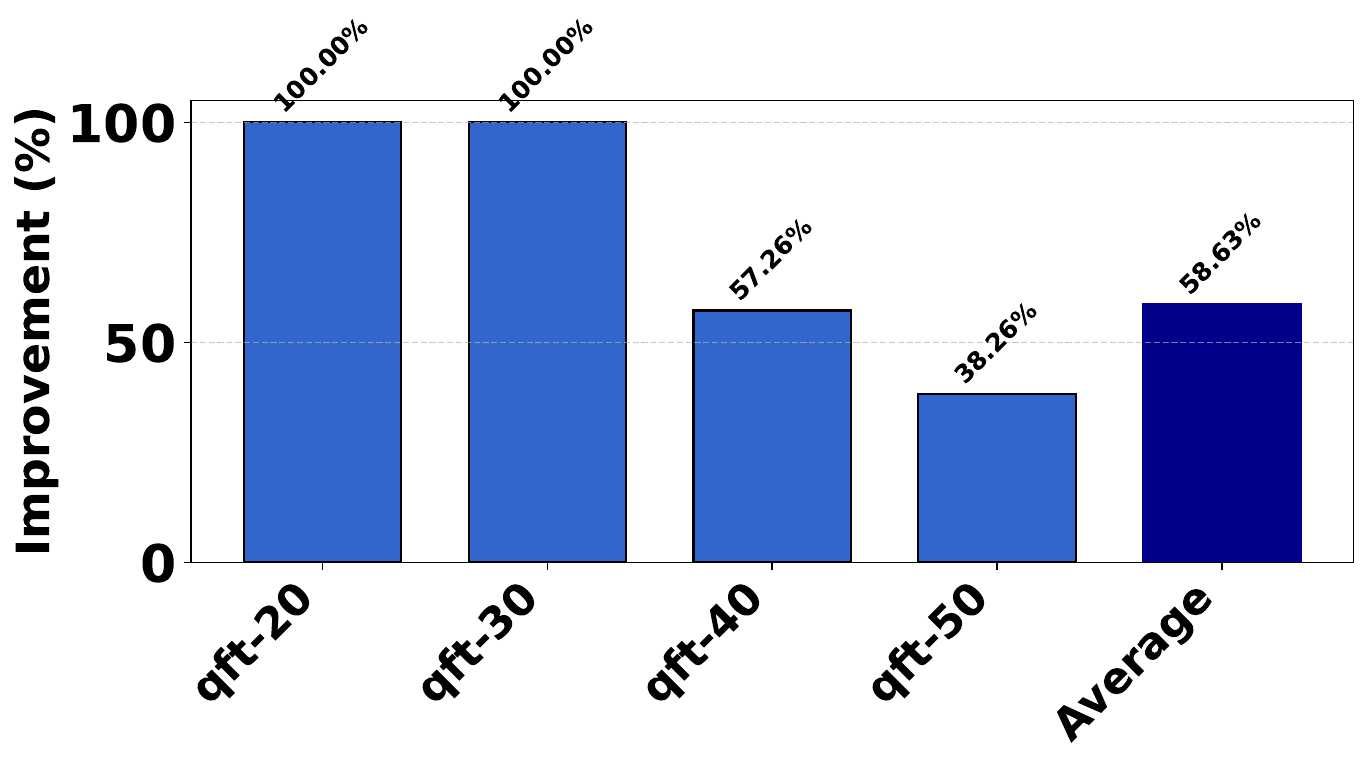}
    \caption{ICCS reduction for Type-I DQCs.}
    \vspace{-5pt}
    \label{fig:type_i_improvement_percentage}
\end{figure}

To the best of our knowledge, no prior work has addressed the same problem as the one we tackle in this study.
All existing layout synthesizers fall back to generate a randomized layout for Type-I DQCs.
In contrast, \name{} is explicitly designed to account for controller-architectural constraints, aiming to minimize \# ICCS.
As shown in Table~\ref{tab:main_res}, \name{} achieves a significant reduction in \# ICCS while keeping the number of operations and circuit depth unchanged.
Notably, for QFT circuits with 20 and 30 qubits, \name{} achieves 100\% reduction of \# ICCS, as the number of qubits is small enough to be mapped entirely within a region managed by a single controller.
As shown in Fig.~\ref{fig:type_i_improvement_percentage}, for QFT circuits with 40 and 50 qubits, \name{} reduces \# ICCS by 57.26\% and 38.26\%, respectively.

\subsection{Performance on Type-II DQCs}

\begin{figure}
    \centering
    \includegraphics[width=0.7\linewidth]{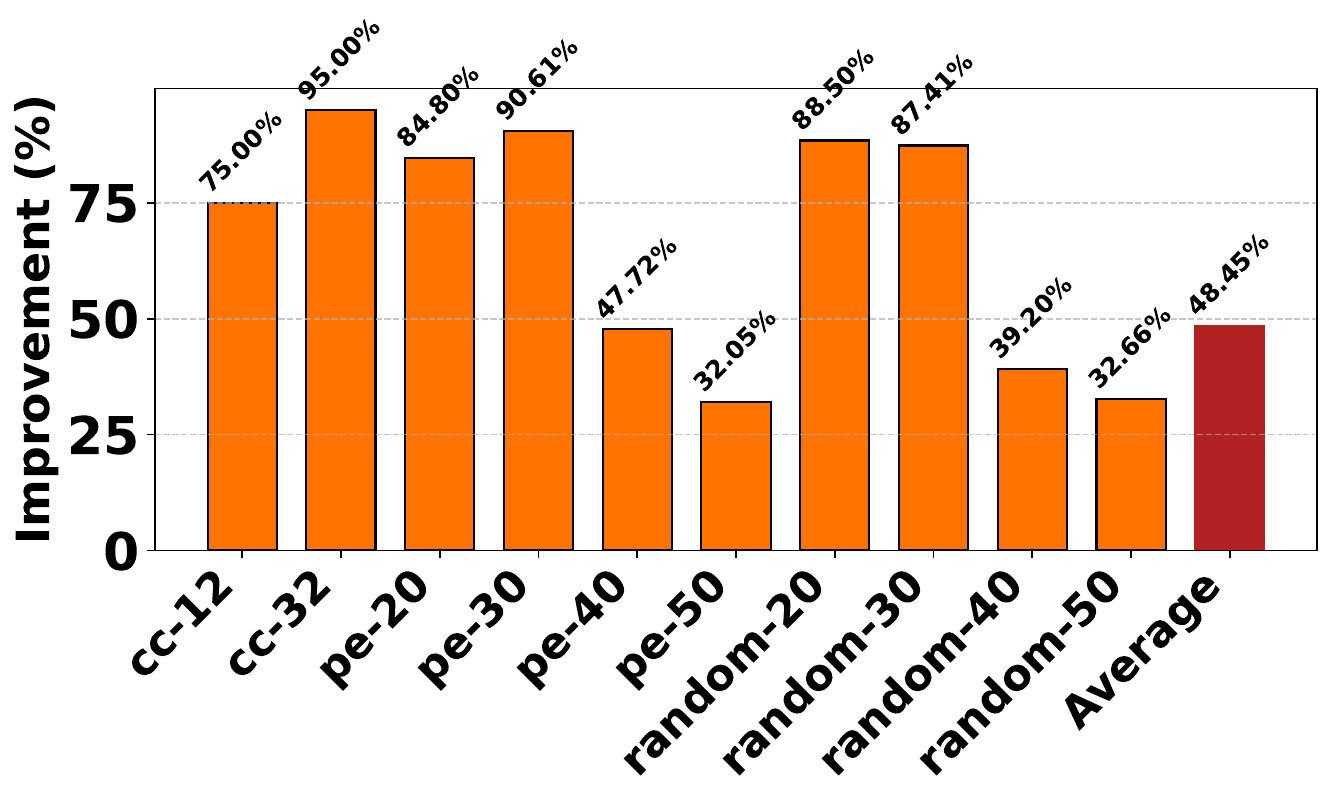}
    \caption{ICCS reduction for Type-II DQCs.}
    \vspace{-5pt}
    \label{fig:type_ii_improvement_percentage}
\end{figure}

For DQCs with connectivity constraints, we also achieve considerable reductions in inter-controller communication latency,
while introducing only a small overhead in terms of the number operations and circuit depth.
As summarized in Table~\ref{tab:main_res}, the post-compilation circuits of the baseline approach and \name{} contain an average of 4404.9 and 4497.4 operations, respectively.
This indicates that \name{} introduces only a modest overhead of approximately 2.10\% in additional CNOT gates.
In contrast, \name{} reduces the average number of ICCSs from 749.20 to 386.20, representing a substantial reduction of approximately 48.45\%, as shown in Fig.~\ref{fig:type_ii_improvement_percentage}.

\begin{figure}
    \centering
    \includegraphics[width=0.7\linewidth]{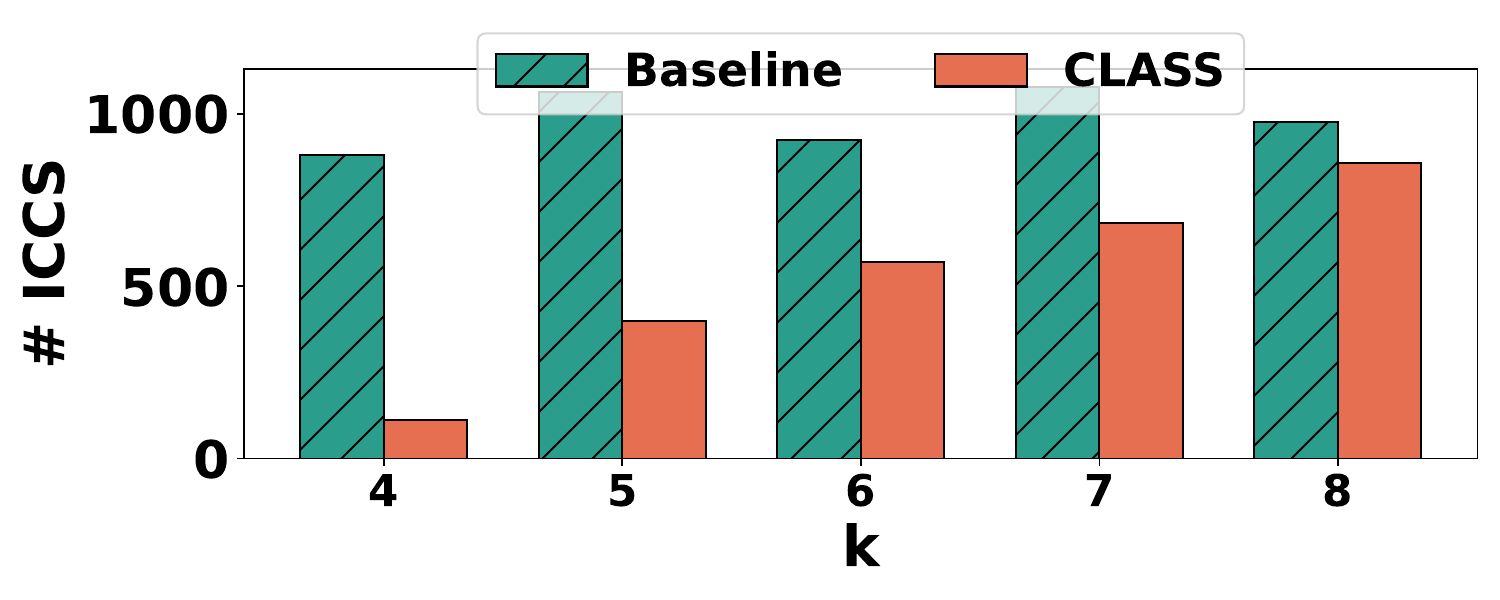}
    \caption{Impact of the number of controllers.}
    \label{fig:impact_ctrl_num}
\end{figure}

\subsection{Impact of Controller Architecture}

We have also verified the adaptability of \name{} to arbitrary controller architecture.
Specifically, we generate various controller topologies with randomized and diverse inter-controller communication latency.
Subsequently, we transpile a 12-qubit QFT circuit onto these architectures and compare the performance between \name{} and baseline.
On average, we obtain 46.71\% reduction in terms of \# ICCS.

Additionally, we conduct a study to evaluate the impact of the number of controllers (denoted as \( k \)) on the performance of \name{}.
For a randomized DQC with 30 qubits, we vary \( k \) from 4 to 8.
As shown in Fig.~\ref{fig:impact_ctrl_num}, the improvement of \name{} over the baseline decreases as \( k \) increases.
This occurs because increasing \( k \) leads to a higher number of subgraphs in the graph-partitioning problem, naturally resulting in more edge cuts between these subgraphs.
In an extreme scenario where each controller manages only a single qubit, inter-controller communications cannot be eliminated, and our approach would show no improvement over the baseline.

\subsection{Scalability Analysis}

\begin{figure}[h!]
    \centering
    \includegraphics[width=0.7\linewidth]{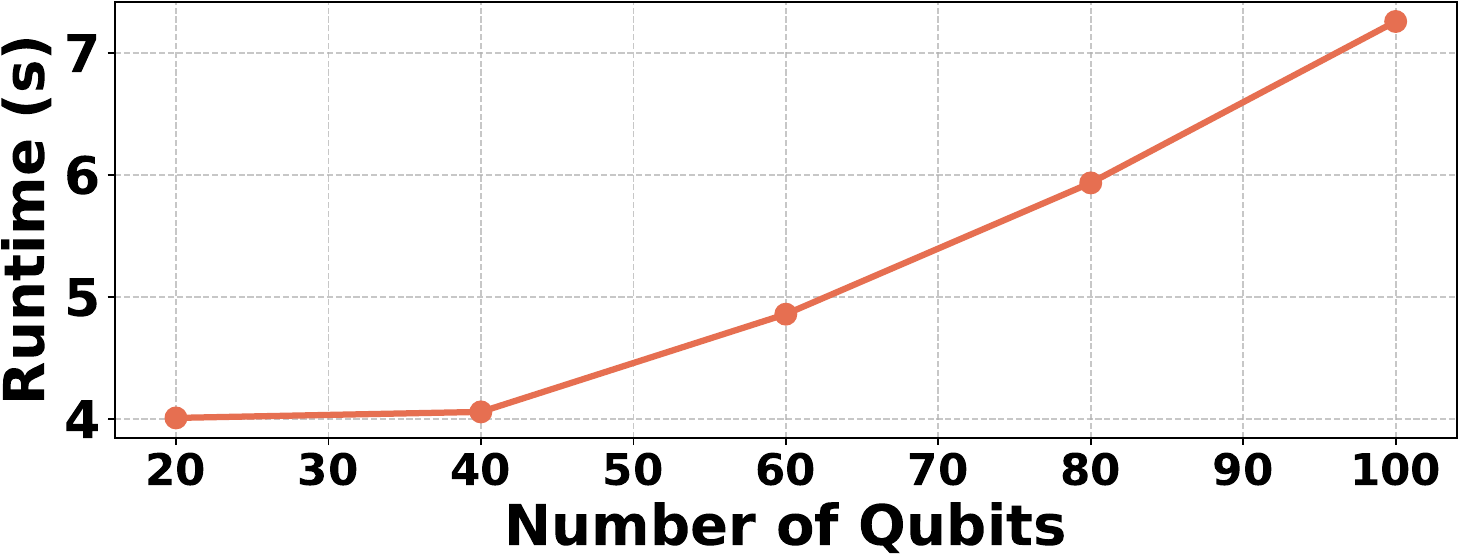}
    \caption{Runtime of ICCS-Aware initial placement vs. \# qubits.}
    \label{fig:runtime_analysis}
\end{figure}

Since our gate scheduler extends existing schedulers, its runtime is primarily influenced by those schedulers.
Thus we focus our evaluation on the performance of our initial placement algorithm.
As analyzed in Sec.~\ref{ssec:type_1_dqc}, the complexity of our algorithm exhibits polynomial growth with respect to the number of qubits and controllers.
In practical quantum computing systems, the number of controllers is typically fixed.
Therefore, we set \( k = 5 \) and vary the number of qubits in dynamic QFT circuits to profile the runtime of our initial placement algorithm.
As shown in Fig.~\ref{fig:runtime_analysis}, the runtime increases from approximately 4 seconds to 7 seconds as the number of qubits grows from 20 to 100.
The efficiency of our algorithm could be further enhanced by employing more efficient system-level programming languages.

%% file: paper/5-discussion.tex
\section{Discussion}

\subsection{Feasibility of \name{} in Quantum Error Correction}

The feasibility of applying \name{} in future QEC scenarios depends on whether topological constraints persist in next-generation devices.
Since many QEC protocols, such as surface codes~\cite{fowler2012surface}, are inherently designed to align with device topology, SWAP-based qubit routing may no longer pose a major challenge.
In this context, \name{} may not be directly applicable to QEC applications.
Nevertheless, we believe the controller-centric design methodology remains relevant due to the continued importance of feedforward operations.

\subsection{Feasibility of \name{} in Real Systems}

It is currently challenging to deploy and evaluate \name{} on real quantum systems.
\name{} requires a distributed quantum control architecture that supports arbitrary feedforward operations, but the design of such systems remains an active research area~\cite{ali_cl_arch_zhang2023classical,qubic2_qce}.
Although industrial platforms such as IBM Quantum Cloud can execute arbitrary DQCs, their control systems are not publicly accessible.
As a result, evaluating the performance of \name{} on public quantum cloud platforms is currently infeasible.
To overcome this limitation, developing an in-house control system becomes necessary—an effort that is currently underway.